\documentclass[aps,amsfonts,pra,twocolumn,showpacs]{revtex4}
\usepackage{epsfig,amsmath,amssymb,bm,epsf,graphics,psfrag}

\def\ket#1{\vert#1\rangle}

\def\ipr#1#2{\langle#1\vert#2\rangle}

\def\Longarrow{\protect\@lra}
\def\@lra{\relbar\joinrel\relbar\joinrel\relbar\joinrel%
          \relbar\joinrel\rightarrow}

\begin{document}
\title{Global entanglement and quantum criticality in spin chains}

\author{Tzu-Chieh Wei}
\affiliation{Department of Physics, 
 University of Illinois at Urbana-Champaign, 
 1110 West Green Street, Urbana, Illinois 61801-3080, U.S.A.}
\author{Dyutiman Das}
\affiliation{Department of Physics, 
 University of Illinois at Urbana-Champaign, 
 1110 West Green Street, Urbana, Illinois 61801-3080, U.S.A.}
 \author{Swagatam Mukhopadyay}
\affiliation{Department of Physics, 
 University of Illinois at Urbana-Champaign, 
 1110 West Green Street, Urbana, Illinois 61801-3080, U.S.A.}
 \author{Smitha Vishveshwara}
\affiliation{Department of Physics, 
 University of Illinois at Urbana-Champaign, 
 1110 West Green Street, Urbana, Illinois 61801-3080, U.S.A.}
 \author{Paul M. Goldbart}
\affiliation{Department of Physics, 
 University of Illinois at Urbana-Champaign, 
 1110 West Green Street, Urbana, Illinois 61801-3080, U.S.A.}
\date{May 26, 2004}
%  \date{\today}

\begin{abstract}
Entanglement in quantum XY spin chains of arbitrary length is investigated 
via a  recently-developed {\it global\/} measure suitable for generic quantum 
many-body systems. 
The entanglement surface is determined over the phase diagram, and found to 
exhibit  structure richer than expected. 
Near the critical line, the entanglement is peaked 
(albeit analytically), consistent with the notion that entanglement---the 
non-factorization of wave functions---reflects quantum correlations. 
Singularity does, however, accompany the critical line, as revealed by 
the divergence of the field-derivative of the entanglement along the line.  
The form 
of this singularity is dictated by the universality class controlling the 
quantum phase transition.
\end{abstract}
\pacs{% 
03.67.Mn, 
% Entanglement production, characterization and manipulation
03.65.Ud, 
% Entanglement and quantum nonlocality 
% (e.g. EPR paradox, Bell's inequalities, GHZ states, etc.)
73.43.Nq, 
% Quantum phase transitions
05.70.Jk% 
% Critical point phenomena
}
\maketitle

\noindent
{\it Introduction\/}---%
Quantum entanglement, a term coined by Schr\"odinger,
has been recognized over the past decade as the central actor in many quantum 
information processing tasks, such as
teleportation~\cite{BennettBrassardCrepeauJozsaPeresWootters93}, 
dense coding~\cite{BennettWiesner92}, 
secure key distribution~\cite{Ekert91}, and perhaps quantum
computation~\cite{NielsenChuang00}.  More recently, entanglement has emerged as 
an actor on the nearby stage of quantum many-body physics, especially for systems 
that exhibit quantum phase transitions~\cite{OsborneNielsen02,OsterlohAmicoFalciFazio02,VidalLatorreRicoKitaev03,SommaOrtizBarnumKnillViola04}, where it can play the role of a diagnostic of quantum correlations.  Quantum phase 
transitions~\cite{Sachdev} are transitions between qualitatively distinct phases 
of quantum many-body systems, driven by quantum fluctuations.  In view of the 
connection between entanglement and quantum correlations, one anticipates that 
entanglement will furnish a dramatic signature of the quantum critical point. 
From the viewpoint of quantum information, the more entangled a state, the more 
resources it is likely to possess.  It is thus desirable to study and quantify 
the degree of entanglement near quantum phase transitions.  By employing 
entanglement to diagnose many-body quantum states one may
obtain fresh insight into the 
quantum many-body problem.  

To date, progress in quantifying entanglement has taken place primarily in 
the domain of bipartite systems~\cite{Horodecki01}.  Much of the previous 
work on entanglement in quantum phase transitions has been based on bipartite 
measures, i.e., focus has been on entanglement either between pairs of 
parties~\cite{OsborneNielsen02,OsterlohAmicoFalciFazio02} or between a part and the 
remainder of a system~\cite{VidalLatorreRicoKitaev03}.  For multipartite systems, 
however, the complete characterization of entanglement requires the consideration of 
multipartite entanglement, for which a consensus measure has not yet emerged. 

Singular and scaling behavior of entanglement near quantum critical points 
was discovered in important work by Osterloh and 
co-workers~\cite{OsterlohAmicoFalciFazio02}, who invoked Wootters' 
{\it bipartite\/} concurrence~\cite{Wootters98} 
as a measure of entanglement.  In the present 
letter, we apply a recently-developed {\it global\/} measure that provides 
a {\it holistic\/}, rather than bipartite, characterization of the 
entanglement of quantum many-body systems.  Our focus is on one-dimensional 
spin systems, specifically ones that are exactly solvable and exhibit 
quantum criticality.   For these systems we are able to determine the 
entanglement analytically, and to observe that it varies in a singular 
manner near the quantum critical line.  This supports the view 
that entanglement---the non-factorization of wave 
functions---reflects quantum correlations. Moreover, the boundaries 
between different phases can be detected by the  entanglement.  

\smallskip
\noindent
{\it Global measure of entanglement\/}---%
To introduce a measure for characterizing the {\it global} entanglement,
consider a general, $n$-partite, normalized pure state: 
$|\Psi\rangle=\sum_{p_1\cdots p_n}\Psi_{p_1p_2\cdots p_n}
|e_{p_1}^{(1)}e_{p_2}^{(2)}\cdots e_{p_n}^{(n)}\rangle$.
If the parties are all spin-1/2 then  
each can be taken to have the basis $\{\ket{\!\uparrow},\ket{\!\downarrow}\}$. 
Our scheme for analyzing the entanglement involves considering how well an 
entangled state can be approximated by some unentangled (normalized) state
(e.g.~the state in which every spin points in a definite direction):
$\ket{\Phi}\equiv\mathop{\otimes}_{i=1}^n|\phi^{(i)}\rangle$.
The proximity of $\ket{\Psi}$ to $\ket{\Phi}$ is 
captured by their overlap; the entanglement of $\ket{\Psi}$ is revealed by the 
maximal overlap~\cite{WeiGoldbart03}
\begin{equation}
\label{eq:lambdamax}
\Lambda_{\max}({\Psi})\equiv\max_{\Phi}|\ipr{\Phi}{\Psi}|\,;
\end{equation}
the larger $\Lambda_{\max}$ is, the less entangled is $\ket{\Psi}$. 
(Note that for a product state, $\Lambda_{\max}$ is unity.) 
If the entangled state consists of two separate entangled pairs
of subsystems,
$\Lambda_{\max}$ is the product of the maximal
overlaps of the two. Hence, it makes sense to quantify
the entanglement of $\ket{\Psi}$ 
via the following {\it extensive\/} quantity~\cite{foot:REE}
\begin{equation}
E_{\log_2}({\Psi})\equiv-\log_2\Lambda^2_{\max}(\Psi),
\label{eq:Entrelate}
\end{equation}
This normalizes to unity the entanglement of EPR-Bell and $N$-party GHZ states, 
as well as gives zero for unentangled states. 
Finite-$N$ entanglement is interesting in the context of 
quantum information processing. To characterize the properties of the 
quantum critical point we use the thermodynamic quantity ${\cal E}$ defined
by
\begin{equation}
{\cal E}\equiv\lim_{N\to\infty}{\cal E}_{N}, \ \ {\cal E}_{N}\equiv
{N}^{-1}E_{\log_2}(\Psi),
\end{equation}
where ${\cal E}_{N}$ is the {\it entanglement density\/}, i.e., the entanglement per particle.  

\begin{figure}
\psfrag{h}{$h$}
\psfrag{r}{$r$}
{\psfrag{E}{${\cal E}_{N} \ $}
\centerline{\psfig{figure=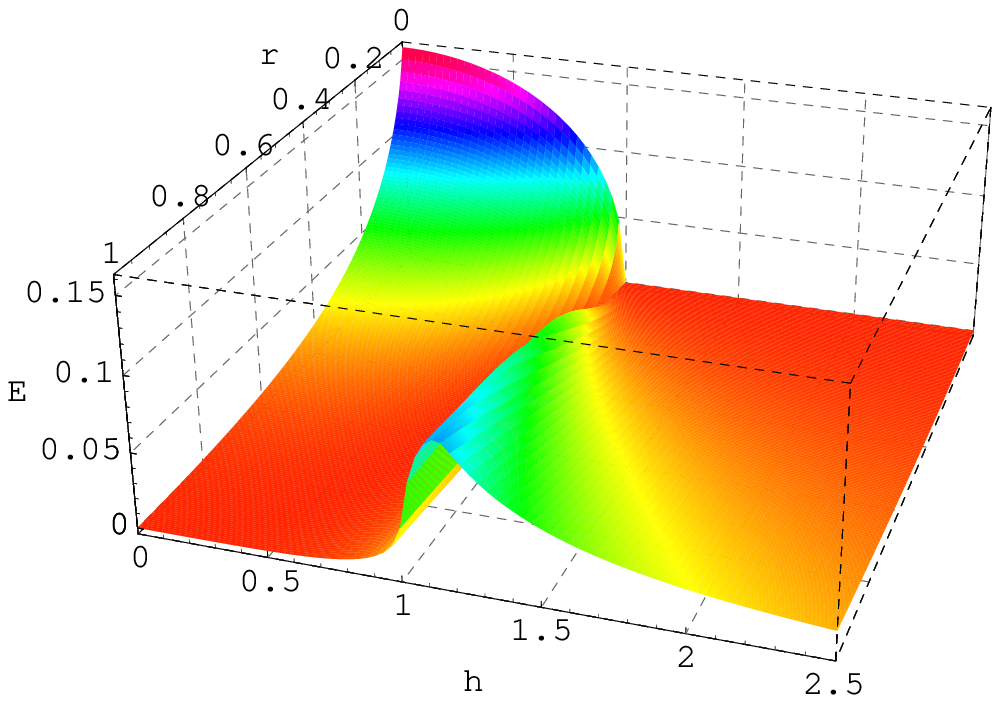,width=8cm,angle=0}}}
%-------------------
% \vspace{-0.4cm}
% \centerline{\psfig{figure=XYentCT.eps,width=7cm,angle=-10}}
\vspace{0.1cm}
  \centerline{\psfig{figure=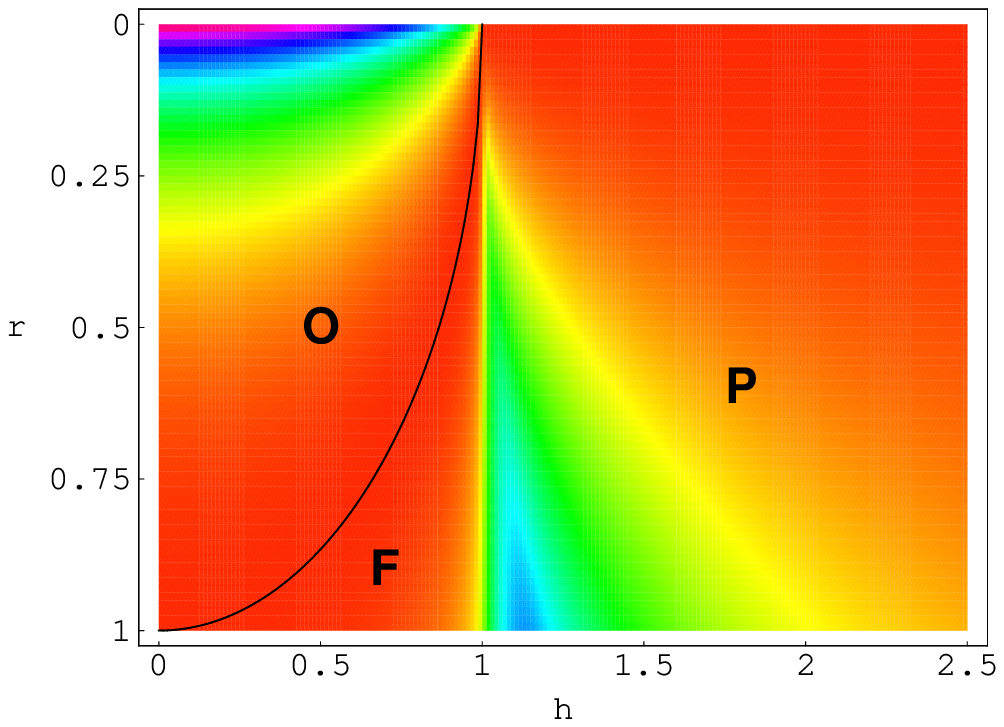,width=7cm,angle=0}}
%-------------------
\caption{(Color online) 
Entanglement density (upper) and phase diagram (lower) 
vs.\ $(r,h)$ for the XY model with $N=10^4$ spins, which
is essentially in the thermodynamic limit.  
There are three phases:
{\bf O}: ordered oscillatory, for $r^2+h^2<1$ and $r\ne 0$; 
{\bf F}: ordered ferromagnetic, between $r^2+h^2>1$ and $h<1$; 
{\bf P}: paramagnetic, for $h>1$. As is apparent, there is a sharp 
rise in the entanglement
across the line $h=1$, which signifies a quantum phase transition. The
arc $h^2+r^2=1$, along which the entanglement density is zero (see also
Fig.~\ref{fig:Ent1000}), separates phases
{\bf O} and {\bf F}. Along $r=0$ lies the XX model, which belongs to a different universality
class from the anisotropic XY model. } 
\label{fig:XYEnt10000}
\end{figure}

\smallskip
\noindent
{\it Quantum XY spin chains and entanglement\/}---% 
We consider the family of models governed by the Hamiltonian 
\begin{equation}
\label{eqn:HXY}
{{\cal H}_{\rm XY}}=
- \sum_{j=1}^N \left(\frac{1\!+\!r}{2}
\sigma_j^x\sigma_{j\!+\!1}^x+
\frac{1\!-\!r}{2}\sigma_j^y\sigma_{j\!+\!1}^y+
h \,\sigma_j^z\right),
\end{equation}
where $r$ measures the anisotropy between $x$ and $y$ couplings, 
$h$ is the transverse external field, lying along the $z$-direction, 
and we impose periodic boundary conditions.
At $r=0$ we have the isotropic XY limit (also known as the XX model)
and 
at $r=1$, the Ising limit. All anisotropic XY models 
($0<r\le 1$) belong to the same universality class, i.e., the Ising 
class, whereas the isotropic XX model belongs to a different universality
class. 
XY models exhibit three phases (see Fig.~\ref{fig:XYEnt10000}): 
oscillatory, ferromagnetic and paramagnetic.
In contrast to the paramagnetic phase, the first two are ordered phases, 
with the oscillatory phase being associated
with a characterstic wavevector, reflecting the modulation
of the spin correlation functions (see, e.g., Ref.~\cite{Henkel99}). 
We shall see that the global 
entanglement detects the boundaries between these phases.

As is well known~\cite{Sachdev,LiebSchultzMattis61,Henkel99}, the energy eigenproblem for the XY spin chain can be solved by a 
Jordan-Wigner transformation, via which the spins are recast as fermions, 
followed by a Bogoliubov transformation,
which diagonalizes the quadratic Hamiltonian.
Having found the eigenstates, $\Lambda_{\rm max}$ of Eq.~(\ref{eq:lambdamax}) 
and hence the entanglement can be found.  
To do this, we parametrize the separable states via
\begin{equation}
\ket{\Phi}\equiv\mathop{\otimes}_{i=1}^{N}
\big[\sin(\xi_i/2)\ket{\!\!\uparrow}_i + 
e^{i\phi_i}\cos(\xi_i/2)\ket{\!\!\downarrow}_i\big],
\end{equation}
where $\ket{\!\uparrow\!\!/\!\!\downarrow}$ denote spin states parallel/antiparallel 
to the $z$-axis.  Instead of maximizing the overlap with respect to the $2N$ real 
parameters $\{\xi_{i},\phi_{i}\}$, for the lowest two states it is
adequate to appeal to the translational symmetry and reality of the wave 
functions.  Thus taking $\xi_i=\xi$ and $\phi_i=0$ we make the Ansatz:
% \begin{equation}
% \label{eqn:PhiTh}
$ \ket{\Phi(\xi)}
\equiv
e^{-i\frac{\xi}{2}\sum_{j=1}^N\sigma_j^y}
\ket{\!\uparrow\uparrow\dots\uparrow}$
%\end{equation}
for searching the maximal the overlap $\Lambda_{\max}(\Psi)$~\cite{footnote:ansatz}.  
This form shows that this separable state can be constructed 
as a global rotation of the ground state at $h=\infty$, viz., the separable 
state $\ket{\!\uparrow\uparrow\dots\uparrow}$. 
The energy eigenstates are readily expressed in terms of the Jordan-Wigner 
fermion operators, and so too are the Ansatz states $\ket{\Phi(\xi)}$.  By 
working in this fermion basis we are able to evaluate the overlaps between 
the two lowest states and the Ansatz states. With $\ket{\Psi_0}$ ($\ket{\Psi_1}$) denoting the lowest state in 
the even (odd) fermion-number sector, we arrive at the overlaps
%---------------
\def\myaa{a}
\begin{widetext}
\begin{equation}
\label{eqn:overlap}
\ipr{\Psi_{\myaa}(r,h)}{\Phi(\xi)}
= 
f^{(\myaa)}_{N}(\xi)\prod_{m=\myaa}^{m<\frac{N-1}{2}}
\left[
\cos\theta^{(\myaa)}_{m}(r,h)\cos^2({\xi}/{2})+
\sin\theta^{(\myaa)}_{m}(r,h)\sin^2({\xi}/{2})
\cot(k_{m,N}^{(\myaa)}/2)
\right], 
\end{equation}
\end{widetext}
\begin{subequations}
% \label{eqn:f}
\begin{eqnarray}
&& k_{m,N}^{(\myaa)}
\equiv
\frac{2\pi}{N}(m+\frac{\myaa}{2}),
\nonumber
\\
&& \tan2\theta^{(\myaa)}_{m}(r,h)
\equiv
r\sin k_{m,N}^{(\myaa)}\big/
(h\!-\!\cos k_{m,N}^{(\myaa)}); 
\nonumber
\\
&& 
\!\!\!\!\!\!\!
f^{(0)}_{N}(\xi)
\equiv 1, 
\ 
f^{(1)}_{N}(\xi)
\equiv
\sqrt{N}\sin({\xi}/{2})\cos({\xi}/{2}), 
\ 
(\mbox{$N$ even}); 
\nonumber
\\ 
&& 
\!\!\!\!\!\!\!
f^{(0)}_{N}(\xi)
\equiv
\cos({\xi}/{2}), 
\ 
f^{(1)}_{N}(\xi)
\equiv
\sqrt{N}\sin({\xi}/{2}), 
\ \ \,\,
(\mbox{$N$  odd}); 
\nonumber
\end{eqnarray}
\end{subequations}
%---------------
where $\myaa=0,1$ and $m\in[0,N-1]$ is an integer.  The above results 
are exact for arbitrary $N$, obtained with periodic boundary conditions on spins rather
than the so-called $c$-cyclic approximation~\cite{LiebSchultzMattis61}.  
Given these overlaps, we can readily obtain the entanglement of 
the ground state, the first excited state, and any linear superposition, 
$\cos\alpha\ket{\Psi_0}+\sin\alpha\ket{\Psi_1}$
of the two lowest states, for arbitrary $(r,h)$ and $N$, 
by maximizing the magnitude of the overlap
with respect to the single real parameter $\xi$.

\begin{figure}
\psfrag{h}{$h$}\psfrag{E}{${\cal E}(h)$}\psfrag{D}{${\cal E}'(h)$}
\centerline{\psfig{figure=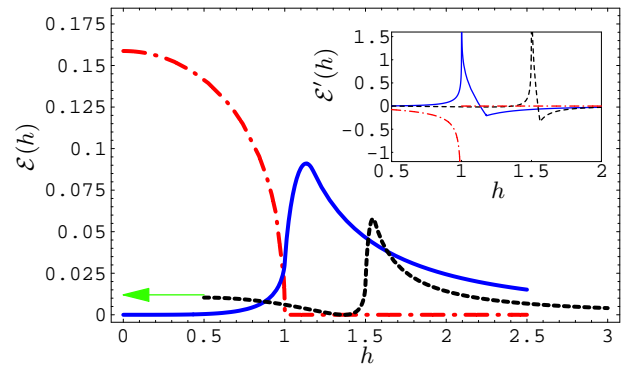,width=8cm,angle=0}}
\caption{(Color online) 
Entanglement density and 
its $h$-derivative (inset)
for the ground state of three systems at $N=\infty$. 
Solid blue line: Ising ($r=1$) limit; 
dashed black line: anisotropic ($r=1/2$) XY model; 
dash-dotted red line: ($r=0$) XX model. 
For the sake of clarity, the XY-case curves are shifted 
to the right by 0.5, indicated by the green arrow. For the $r=1/2$, at $h^2+r^2=1$ the entanglement
density vanishes, which is a general property for the anisotropic XY
model.
Note that whilst the the entanglement itself has a nonsingular 
maximum at $h\approx 1.1$ (Ising), $h\approx 1.04$ (XY $r=1/2$), 
$h=0$ (XX), respectively,
it has a singularity at the quantum critical point at $h=1$, 
as revealed by the divergence of its derivative.} 
\label{fig:Ent1000}
\end{figure}

The formulas that we have just established contain all the results 
that we explore in the present letter. By analyzing the
structure of Eq.~(\ref{eqn:overlap}), we find that the global
entanglement does provide information on the phase structure
and critical properties of the quantum spin chains.  Two of our key results, 
as captured 
in Figs.~\ref{fig:XYEnt10000} and \ref{fig:Ent1000},
are that: (i)~although the entanglement itself is, generically, 
not maximized at the quantum critical line in the 
$(r,h)$ plane, {\it the field-derivative of the entanglement 
diverges as the critical line $h=1$ is approached\/}; and (ii)~the entanglement {\it vanishes\/} at the disorder line $r^2+h^2=1$, which 
separates the oscillatory and ferromagnetic phases. 

\smallskip
\noindent
{\it Entanglement and quantum criticality\/}---% 
From Eq.~(\ref{eqn:overlap}) it follows that the thermodynamic 
limit of the entanglement density is given by
\begin{eqnarray}
{\cal E}(r,h)
&=&
-\frac{2}{\ln2}\max_\xi
\int_0^{\frac{1}{2}}d\mu\,
\ln\left[\cos\theta(\mu,r,h)\cos^2({\xi}/{2})\right.
\nonumber\\
&&\left. 
\qquad\quad
+\sin\theta(\mu,r,h)\sin^2({\xi}/{2})\cot\pi\mu\right],
\label{eq:infNent}
\end{eqnarray}
where $\tan 2\theta(\mu,r,h)\equiv r\sin 2\pi\mu /(h-\cos 2\pi\mu)$. 

Figure~\ref{fig:Ent1000} shows the thermodynamic limit of the 
entanglement density ${\cal E}(r,h)$ and its $h$-derivative in 
the ground state, as a function of $h$ for three values of $r$, 
i.e., three slices through the surface shown in 
Fig.~\ref{fig:XYEnt10000}.  
As the $r=1$ slice shows, in the Ising 
limit the entanglement density is small for both small and large $h$.  
It increases with $h$ from zero, monotonically, albeit very slowly for 
small $h$, then swiftly rising to a maximum at $h\approx 1.13$ 
before decreasing monotonically upon further increase of $h$, 
asymptotically to zero.  The entanglement maximum {\it does not\/} occur 
at the quantum critical point.  However, the derivative of 
the entanglement with respect to $h$ {\it does\/} diverge at the critical 
point $h=1$, as shown in the inset.  
The slice at $r=1/2$ (shifted, 
for clarity, half a unit to the right) shows qualitatively similar 
behavior, except that it is finite (although small) at $h=0$, and 
starts out by decreasing to a shallow minimum of zero at 
$h=\sqrt{1-r^{2}}$.  
By constrast, the slice at $r=0$ (XX) starts out at $h=0$ at a 
maximum value of  $1- 2 \gamma_C/(\pi \ln 2)\approx 0.159$. 
(where $\gamma_C$ is the {\it Catalan\/} constant), the globally 
maximal value of the entanglement over the entire $(r,h)$ plane. 
% Catalan is $\approx 0.9160$
For larger $h$ it falls monotonically until it vanishes at $h=1$, 
remaining zero for larger $h$.  

We find that along the line $r^{2}+h^{2}=1$ the entanglement 
density vanishes in the thermodynamic limit. In fact, 
this line exactly corresponds to
the boundary separating the oscillatory and ferromagnetic phases;
the boundary can be characterized by a set of ground
states with total entanglement of order unity, and thus
of zero entanglement density. The entanglement density is
also able to track the phase boundary ($h=1$) between the ordered
and disordered phases. Associated with the quantum fluctuations
accompanying the transition, the entanglement density shows
a drastic variation across the boundary and the field-derivative
diverges all along $h=1$. The two boundaries separating the three
phases coalesce at $(r,h)=(0,1)$, i.e., the XX critical point. 
Figures~\ref{fig:XYEnt10000} and \ref{fig:Ent1000} reveal all
these features.

The singular behavior of the entanglement density~(\ref{eq:infNent}) can be 
analyzed in the vicinity of the quantum critical line, and 
we find the asymptotic behavior (for $r\ne 0$)
\begin{equation}
\label{eqn:EntDiv}
\frac{\partial{\cal E}}{\partial h}
\approx 
-\frac{1}{2\pi r \ln2}\ln|h-1|,
\,\,\,
\mbox{for $\vert{h-1}\vert\ll 1$}.
\end{equation}
From the arbitrary-$N$ results~(\ref{eqn:overlap}) we analyze the approach 
to the thermodynamic limit, in order 
to further understand connections with quantum criticality. 
We focus on the exponent $\nu$, which governs the divergence at 
criticality of the correlation length: $L_c\sim |h-1|^{-\nu}$.
To do this, we compare the divergence of the slope 
$\partial{\cal E}_{N}/\partial h$
(i)~near $h=1$ (at $N=\infty$), given above, and 
(ii)~for large $N$ at the value of $h$    
for which the slope is maximal (viz.\ $h_{{\rm max},N}$), i.e., 
$\partial{\cal E}_{N}/\partial h
\vert_{h_{{\rm max},N}}\approx
0.230r^{-1}\ln N + {\rm const.}$, 
 obtained by analyzing Eq.~(\ref{eqn:overlap})
for various values of $r$.
Then, noting that $(2\pi\ln2)^{-1}\approx 0.2296$ 
and that the logarithmic scaling hypothesis~\cite{Barber83} 
identifies $\nu$ with the ratio of the amplitudes of these divergences,  
$0.2296/0.230\approx 1$, we recover the known result that $\nu=1$.

Compared with $r\ne 0$ case, the nature of the divergence of 
$\partial{\cal E}/\partial h$ at $r=0$  belongs to a different
universality class:
\begin{equation}
\label{eqn:EntDivXX}
\frac{\partial}{\partial h}{\cal E}(0,h)\approx
-\frac{\log_2(\pi/2)}{\sqrt{2}\,\pi}
\frac{1}{\sqrt{1-h}},
% (1-h)^{-\frac{1}{2}}, 
\qquad (h\to 1^{-}). 
\end{equation}
From this divergence, the scaling hypothesis, and the assumption that
the entanglement density is intensive,  
we can infer the known result that the critical 
exponent $\nu=1/2$ for the XX model.
In keeping with the critical features of the XY-model phase diagram,
for any small but nonzero value of the anisotropy, 
the critical divergence of the entanglement derivative 
is governed by Ising-type behavior. 
It is only at the $r=0$ point that the critical behavior of
the entanglement is governed by the XX universality class.
For small $r$, XX behavior ultimately crosses over to Ising behavior.

As is to be expected, at finite $N$ the two lowest states 
$\ket{\Psi_0}$ and $\ket{\Psi_1}$ featuring in Eq.~(\ref{eqn:overlap})
do not spontaneously break the ${\rm Z}_2$ symmetry. However, in the
thermodynamic limit they are degenerate for $h\le 1$, 
and linear combinations are also ground states. The question then arises
as to whether linear combinations that explicitly break ${\rm Z}_2$
symmetry, i.e., the physically relevent states with finite spontaneous
magnetization, show the same entanglement properties. In fact, we see
from Eq.~(\ref{eqn:overlap}) that, in the thermodynamic limit, overlaps
for both $\ket{\Psi_0}$ and $\ket{\Psi_1}$ are identical, up to the prefactors 
$f^{(0)}_N$ and $f^{(1)}_N$. These prefactors do not contribute to the
entanglement density, and the entanglement density
is therefore the same for both $\ket{\Psi_0}$ and $\ket{\Psi_1}$. 
It  follows that,
in the thermodynamic limit, the results for the entanglement density 
are insensitive to the replacement of a symmetric ground state by a 
broken-symmetry one.

\smallskip
\noindent
{\it Concluding remarks\/}---% 
In summary, we have quantified the global entanglement of the quantum 
XY spin chain.  This model exhibits a rich phase structure, the 
qualitative features of which are reflected by this entanglement measure. 
Perhaps the most interesting aspect is the divergence of the 
field-derivative of the entanglement as the critical 
line ($h=1$) is crossed.  
Furthermore, in the thermodynamic
limit, the entanglement density vanishes on the disorder line ($r^2+h^2=1$). 
The structure of the 
entanglement surface, as a function of the parameters of the model 
(the magnetic field $h$ and the coupling anistotropy $r$), is 
surprisingly rich. 

We close by pointing towards a deeper connection 
between the global measure of entanglement and the correlations 
among quantum fluctuations. The maximal overlap~(\ref{eq:lambdamax})
can be decomposed in terms of correlation functions:
\begin{equation}
\Lambda_{\max}^2
=\frac{1}{2^N}+
\max_{|\vec{r}|=1}
\frac{N}{2^N}\Big\{\langle\vec{r}\cdot\vec{\sigma}_{1}\rangle+
\sum_{j=2}^N\langle\vec{r}\cdot \vec{\sigma}_{1}
\otimes\vec{r}\cdot \vec{\sigma}_{j}\rangle+
\cdots\Big\},\nonumber
\end{equation}
where 
translational invariance is assumed and the Carteisan coordinates of $\vec{r}$ can be taken to be 
$(\sin\xi,0,\cos\xi)$. The two-point correlations appearing in the
decomposition are related to a bipartite measure of entanglement, namely, the 
concurrence, which shows similar singular behavior~\cite{OsterlohAmicoFalciFazio02}. It would be interesting to establish
the connection between the global entanglement and correlations more precisely, e.g., by identifying which correlators are responsible for the singular behavior
in the entanglement and how they relate to the better known critical
properties. 

\smallskip
\noindent
{\it Acknowledgments\/}---%
We thank A.~Abanov, J.-S.~Caux, D.~Chang, E.~Kim, P.~Kwiat, 
G.~Ortiz, M.~Randeria, M.~Stone, N.~Trivedi,  
W. Zhang, and especially E.~Fradkin and A.A.~Ludwig 
for valuable discussions.
This work was supported by 
NSF EIA01-21568 and DOE DEFG02-96ER45434.


\begin{thebibliography}{99}
\bibitem{footnote:ansatz}
We have confirmed this Ansatz numerically for small numbers of spins.
\end{thebibliography}
\end{document}